\documentclass[letterpaper, 10 pt, conference]{ieeeconf}  

\IEEEoverridecommandlockouts                              

\usepackage{graphics} 
\usepackage{epsfig} 
\usepackage{times} 

\usepackage{amsthm}
\usepackage{amsmath} 
\usepackage{amssymb}  
\usepackage{mathtools}
\usepackage[short]{optidef}
\usepackage{tensor}
\usepackage{units}
\usepackage{tikz}
\usetikzlibrary{decorations.pathreplacing,calc}
\usepackage{pgfplots}
\pgfplotsset{compat=newest}
\usepackage{url}

\usepackage{relsize}
\usepackage{booktabs}
\usepackage{tikz}
\usetikzlibrary{calc,patterns,angles,quotes}
\usepackage{colortbl}%

\usepackage{eqparbox}
\usepackage{svg}
\usepackage{makecell}
\usepackage{boldline}
\usepackage{cite}

\theoremstyle{plain}
\newtheorem{assumption}{Assumption}

\theoremstyle{plain}
\newtheorem{theorem}{Theorem}

\theoremstyle{plain}
\newtheorem{corollary}{Corollary}

\mathtoolsset{showonlyrefs}

\title{\LARGE \bf
Imitation Learning of MPC with Neural Networks: Error Guarantees and Sparsification
}

\author{Hendrik Alsmeier$^{1}$, Lukas Theiner$^{1}$, Anton Savchenko$^{1}$, Ali Mesbah$^{2}$, Rolf Findeisen$^{1}$
\thanks{*The authors acknowledge funding of the KI-Embedded
project of the German Federal Ministry of Economic Affairs
and Climate Action (BMWK).}
\thanks{$^{1}$Control and Cyber-Physical Systems Laboratory, Technical University of Darmstadt, Germany,\newline \{hendrik.alsmeier, lukas.theiner, anton.savchenko, rolf.findeisen\}@iat.tu-darmstadt.de}%
\thanks{$^{2}$Dept of Chemical \& Biomolecular Engineering, University of California, Berkeley, USA, {mesbah@berkeley.edu}}%
}

\begin{document}
\bstctlcite{IEEEexample:BSTcontrol}

\maketitle
\thispagestyle{empty}
\pagestyle{empty}

\begin{abstract}

This paper presents a framework for bounding the approximation error in imitation model predictive controllers utilizing neural networks. Leveraging the Lipschitz properties of these neural networks, we derive a bound that guides dataset design to ensure the approximation error remains at chosen limits. We discuss how this method can be used to design a stable neural network controller with performance guarantees employing existing robust model predictive control approaches for data generation. Additionally, we introduce a training adjustment, which is based on the sensitivities of the optimization problem and reduces dataset density requirements based on the derived bounds. We verify that the proposed augmentation results in improvements to the network's predictive capabilities and a reduction of the Lipschitz constant. Moreover, on a simulated inverted pendulum problem, we show that the approach results in a closer match of the closed-loop behavior between the imitation and the original model predictive controller.
 
\end{abstract}

\section{INTRODUCTION}
Model predictive control (MPC) has gained significant attention as an approach for controlling a wide range of systems~\cite{schwenzer2021,lucia2016predictive}, as it can handle complex nonlinear systems and effortlessly incorporate input and state constraints into its optimal control problem \cite{rawlings2017,findeisen2002}.
However, as system complexity grows, the computational cost increases, often leading to challenges in real-time capability, as the solution time of the optimal control problem rises \cite{rawlings2017}. This has prompted interest in substituting the resource-intensive MPC with an approximation of its implicit control law to avoid online optimization. Several methods have been developed to generate such approximations, ranging from the pre-calculation of the control law in explicit MPC~\cite{tondel2003algorithm} to learning frameworks for reinforcement learning based on MPC~\cite{zanon2020}.\\
Neural networks (NNs) have proven to be a promising candidate for the approximation of MPC, and several learning algorithms to generate an approximate controller have been introduced~\cite{karg2020,paulson2020,zhang2019,kohler2018,Allgoewer2018a,chen2018,hewing2020,nguyen2021robust,mesbah2022fusion}. 
However, the approximation of MPC also introduces challenges, primarily due to the loss of explicit guarantees in the generated controller, which generally can no longer be explicitly enforced in the approximate controller. A perfect approximation of MPC is generally not possible, leading to approximation errors that can be amplified in a closed-loop setting, where faulty control is fed back into the system. Addressing these errors is crucial, especially in safety-critical applications, where constraint satisfaction is paramount.   
In literature, this challenge is handled by bounding the approximation error. For instance, in \cite{limon2017}, a specialized method is used to bound the approximation error for noisy data.  An approach for neural networks is presented in \cite{drummond2022}, where calculating a bound via optimization is presented for the approximation of MPC with a quadratic program as an optimal control problem. Additionally, in \cite{Allgoewer2018a} a method for fulfilling a set bound via robust MPC for NN approximations is put forward.\\
In this work, we propose a constructive approach to derive the bound and design a NN controller, which is different from most existing approaches, like \cite{Allgoewer2018a}. These approaches often rely on post-verification to ensure that a desired approximation error is not exceeded after the training of the NN. 
The approach leverages the Lipschitz properties of NNs to derive an analytical worst-case bound for the deviation of control inputs generated by the NN approximation and the original MPC law. Furthermore, we provide a way to increase sampling efficiency to fulfill this bound by drawing from recent advances in dataset augmentation of the learned controller by utilizing sensitivities of nonlinear programs \cite{Krishnamoorthy2022,luken2023}.
The remainder of this paper is structured as follows. In Section \ref{sec:probsetup}, we give further details of our considered setting, as well as briefly introduce the concept of MPC and the desired approximation. Then, in Section \ref{sec:perfguar} we discuss the derivation of the approximation error bound and show how a robust MPC can be used to generate a stable and recursively feasible controller using this bound. After this, in Section \ref{sec:sensreg} we introduce a training modification for the neural network to reduce the needed sample density and give a brief discussion on the calculation of nonlinear program sensitivities. In Section \ref{sec:sim}, we show how our augmentation can enhance NN performance via the closed-loop simulation of learned MPC for an inverted pendulum. We conclude with summarizing remarks and outlooks for future research.

\section{PROBLEM SETUP}
\label{sec:probsetup}
We consider nonlinear systems of the following form, e.g. obtained from discretizing a continuous time system,
\begin{equation}
        x(t_{n+1}) = f(x(t_n),u(t_n)). \label{eq:discdynamics}
\end{equation}
Here, ${x(\cdot) \in \mathcal{X} \subseteq \mathbb{R}^{n_x}}$ denotes the state and ${u(\cdot)\in \mathcal{U} \subseteq\mathbb{R}^{n_u}}$ the input of the system, with dimensions $n_x$ and $n_u$, respectively, and the sampling times $t_n$. For the remainder of this work, we omit the time dependency of $x$ and $u$. ${f: \mathcal{X} \times \mathcal{U} \rightarrow \mathcal{X}}$ defines the state transition for a given input and state.\\
The system \eqref{eq:discdynamics} can be controlled via MPC, where the applied input is given by the repeated solution of the optimal control problem (OCP)
\begin{argmini}[1]
    {\mathbf{u}_k}{\sum^{N-1}_{k=0}S \left(x_k,u_k\right)+V(x_N)}{ \label{eq:mpc}}{\mathrm{OCP}(x_s)=}
    \addConstraint{x_{k+1}}{=f\left(x_k,u_k\right)}{\ \forall k\in [0,\hdots,N\!-\!1]}
    \addConstraint{x_k}{\in\mathcal{X}_k}{\ \forall k\in [0,\hdots,N]}
    \addConstraint{u_k}{\in\mathcal{U}_k}{\ \forall k\in [0,\hdots,N\!-\!1]}
    \addConstraint{x_0}{=x_s}.
\end{argmini}
With $x_k$ and $u_k$, we denote the states and inputs of the predicted trajectory at each time point $k$ in the horizon with the length $N$, to distinguish it from the real trajectory. The stage cost is represented by $S(x_k,u_k)$, while $\mathbf{u}_k$ refers to the sequence of decision variables $[u_0,\ldots,u_{N-1}]$ in the OCP. The constraint sets for states and inputs are given by $\mathcal{X}_k \subseteq \mathcal{X}$ and $\mathcal{U}_k \subseteq \mathcal{U}$. The initial state of the OCP at sampling time $t_s$ is defined as $x_s = x(t_s)$. To apply MPC we need to repeatedly solve \eqref{eq:mpc} . For this, we acquire a sample point from the controlled system $x(t_s)$. Then the OCP is solved, which produces an optimal sequence of inputs $\mathbf{u}_k^{*}=[u_0^*,\hdots,u_{N-1}^*]$, given the stage cost and constraints. The first input $u_{\mathrm{opt}} = u_0^*$ is then applied to the system until the next sampling time $t_{s+1}$ and the process is repeated~\cite{findeisen2002,rawlings2017}.\\
This framework can become computationally expensive, especially for systems with nonlinearities and high state dimensions due to the need to solve a non-convex, nonlinear program at every sampling step the OCP is computed~\cite{rawlings2017}. Therefore, it is often necessary to reduce this complexity by approximating the MPC or more specifically the implicit control law it provides
\begin{equation}
    u_\mathrm{opt} = \kappa(x) := \mathrm{OCP}(x)\big|_{u_0}, \label{eq:mpcmapping}
\end{equation}
where $ u_{\mathrm{opt}} \in \mathcal{U} \subseteq \mathbb{R}^{n_u}$ and $\kappa: \mathcal{X} \rightarrow \mathcal{U}$ denotes the mapping the MPC algorithm performs in the closed-loop. The approximation of \eqref{eq:mpcmapping} can be written as
\begin{equation}
    \tilde u_\mathrm{opt} = \kappa_{\mathrm{a}}(x), \label{eq:nnmapping}
\end{equation}
with $\tilde u_{\mathrm{opt}} \in \tilde{\mathcal{U}} \subseteq \mathbb{R}^{n_u}$ and $\kappa_{\mathrm{a}}: \mathcal{X} \rightarrow \tilde{\mathcal{U}}$.\\
The approximate MPC, in general, loses constraint satisfaction and stability guarantees. The approximation will not be perfect in most cases due to an approximation error at every $x \in \mathcal{X}$. To provide guarantees on the performance of the approximate controller, we therefore aim to determine a bound on the approximation error. Given such a bound, we can establish guarantees for constraint violations and stability when applying the approximate controller. Such bounds can  be used in MPC schemes, which are robust to input uncertainties, to provide closed-loop guarantees, or to adjust the MPC control law to be robust concerning these bounds. Examples of such scenarios are outlined in \cite{Allgoewer2018a} for NNs, or \cite{rose2023learning} for Gaussian processes.

\section{PERFORMANCE GUARANTEES FOR NN APPROXIMATED CONTROLLERS}
\label{sec:perfguar}

We use neural networks to approximate the MPC law $\kappa_{\mathrm{a}}(x) = \kappa_{\mathrm{NN}}(x)$ because they satisfy the universal approximation theorem \cite{HORNIK1991,hornik1993}, enabling them to learn our potentially complex mapping~\eqref{eq:mpcmapping}. For this work, we consider feed-forward NNs, which we train with supervised learning. This architecture comprises neurons into layers, including an input layer, several hidden layers, and an output layer. The value of neuron vector $z_j$ at layer $j$ can be represented as an affine matrix equation with  weight matrix $W_j$, bias vector $b_j$, and the preceding layer's neurons value vector $z_{j-1}$, followed by an element-wise nonlinear activation function $\alpha_j$, expressed as
\begin{equation}
    z_j = \alpha_j\circ\zeta_j(z_{j-1}):= \alpha_j \left( W_{j} z_{j-1} + b_j \right). \label{eq:nnlayer}
\end{equation}
Information flows from the input layer through the $m-1$ hidden layers to the output layer. Therefore, the NN considered for our task can be written as
\begin{equation}
    \tilde u_{\mathrm{opt}} = \kappa_{\mathrm{NN}}(x):= \zeta_m \circ \alpha_{m-1}\circ \zeta_{m-1} \circ \hdots \circ \alpha_1 \circ \zeta_1(x).\label{eq:nndesign}
\end{equation}
In this structure, we have no activation on the output layer $\alpha_m$.
Now we will demonstrate how a worst-case bound of the NN's approximation error when learning a MPC can be derived. To do so, we need the following two assumptions to hold regarding the MPC and the NN training.
\begin{assumption}\label{a:1}
    The MPC control law defined in \eqref{eq:mpcmapping} satisfies the Lipschitz condition, i.e. for $L_{\mathrm{MPC}} > 0$
    \begin{equation}
        \lVert \kappa(x_\mathrm{a}) - \kappa(x_\mathrm{b}) \rVert \leq L_{\mathrm{MPC}} \lVert x_\mathrm{a}- x_\mathrm{b} \rVert,
    \end{equation}
    where $x_\mathrm{a},x_\mathrm{b}$ are arbitrary points in $\mathcal{X}$.
\end{assumption}
Here $\lVert \cdot\rVert$ defines the Euclidean norm and we use $L_\zeta$ to denote the Lipschitz constant of a given operator $\zeta$. This assumption is necessary since we cannot guarantee it to hold for general nonlinear MPC. For linear MPC, the Lipschitz constant can be calculated~\cite{teichrib2023}. 
Feed-forward NNs with commonly used activation functions (\emph{tanh}, \emph{sigmoid}, \emph{ReLU}, etc.) are Lipschitz-continuous by construction \cite{Hastie2009,bhowmick2021}. There exist several techniques to estimate the Lipschitz constant of a neural network \cite{Pappas2019,pauli2021}.
\begin{assumption}\label{a:2}
    For any $\epsilon_\mathrm{D} > 0 $ and a training dataset ${\mathcal{D} = \{(x_i,\kappa(x_i))\}_{i \in \mathcal{I}}}$with an index set I, there exists a sufficiently large NN structure and training procedure such that
    \begin{equation}
        \lVert \kappa(x_i)-\kappa_{\mathrm{NN}}(x_i) \rVert \leq \epsilon_\mathrm{D}~\forall i \in \mathcal{I}, \label{eq:epsd}
    \end{equation}
    for the MPC law $\kappa$ \eqref{eq:mpcmapping} and its NN approximation $\kappa_{\mathrm{NN}}$ \eqref{eq:nnmapping}.
\end{assumption}
This is a mild assumption, as we actively minimize these errors at the training points $x_i$ via the loss function in classical supervised learning. In practice, we can calculate this error as the maximum over the training dataset
\begin{equation}
    \epsilon_\mathrm{D}:= \max_{i \in \mathcal{I}} \lVert \kappa(x_i) - \kappa_\mathrm{NN}(x_i) \rVert, \label{eq:explespd}
\end{equation}
and terminate the training procedure when it reaches the desired value
The size of $\mathcal{D}$ poses no restrictions, as modern NNs can be efficiently trained on large datasets. With these assumptions, we can now derive a global error bound.
\begin{theorem}\label{t:1}
    If Assumptions \ref{a:1} and \ref{a:2} hold, then for any $\epsilon > 0$ the approximation error is bounded by
    \begin{equation}
         \lVert \kappa(x) - \kappa_{\mathrm{NN}}(x) \rVert \leq \epsilon \quad \forall x \in \mathcal{X}, \label{eq:errorbound}
    \end{equation}
    if the following conditions for the dataset $\mathcal{D}$ hold:
    \[\epsilon_D<\epsilon,\quad\delta := \left(L_{\mathrm{MPC}}+L_{\mathrm{NN}}\right)^{-1}\left( \epsilon - \epsilon_\mathrm{D} \right),\]
    for $L_{\mathrm{NN}}$ and $L_\mathrm{MPC}$ being Lipschitz constants of $\kappa_{\mathrm{NN}}$ and $\kappa$ correspondingly, such that
    \begin{equation}
            \forall  x \in \mathcal{X} \quad \exists i \in \mathcal{I}: \lVert x - x_i \rVert \leq \delta \label{eq:delta}.
    \end{equation}
\end{theorem}
This theorem gives guidance on the density of the dataset and can be proven by using Lipschitz constants.

\begin{proof}
    We choose any $x \in \mathcal{X}$ and find the corresponding point $x_i$ such that \eqref{eq:delta} holds. Then the left-hand side of \eqref{eq:errorbound} can be written as
    \begin{equation}
        \begin{split}\lVert \kappa(x)\!-\!\kappa_{\mathrm{NN}}(x) \rVert\!=\lVert &\kappa(x)\!-\!\kappa(x_i)\!+\! \kappa_{\mathrm{NN}}(x_i)-\\&\!\kappa_{\mathrm{NN}}(x)\!+\!\kappa(x_i)\!-\!\kappa_{\mathrm{NN}}(x_i)\rVert\end{split}
    \end{equation}
    Triangle inequality $\lVert a + b \rVert \leq \lVert a \rVert + \lVert b \rVert$ lets us bound it to
    \begin{equation}
        \begin{split}  \lVert \kappa(x)\! -\! \kappa(x_i)\! +\! \kappa_{\mathrm{NN}}(x_i) &-
        \kappa_{\mathrm{NN}}(x)\ +\\ \kappa(x_i)\! -\! \kappa_{\mathrm{NN}}(x_i) \rVert &\leq
        \lVert \kappa(x)\! -\! \kappa(x_i) \rVert \ + \\\lVert \kappa_{\mathrm{NN}}(x_i)\! - \!
        \kappa_{\mathrm{NN}}(x) \rVert & + \lVert  \kappa(x_i)\! -\! \kappa_{\mathrm{NN}}(x_i) \rVert. \end{split} \label{eq:reformbound}
    \end{equation}
    Next, we use Lipschitz conditions from Assumption \ref{a:1}
    \begin{align*}
       \lVert \kappa(x) - \kappa(x_i) \rVert &\leq L_{\mathrm{MPC}} \lVert x - x_i \rVert,\\ 
       \lVert \kappa_{\mathrm{NN}}(x) - \kappa_{\mathrm{NN}}(x_i) \rVert &\leq L_{\mathrm{NN}} \lVert x - x_i \rVert, 
    \end{align*}
     with $L_{\mathrm{MPC}},L_{\mathrm{NN}} > 0$ being the Lipschitz constants, and Assumption~\ref{a:2}, to further bound \eqref{eq:reformbound} by
    \begin{equation*}
        \begin{split}\lVert \kappa(x) - \kappa(x_i) \rVert  &+ \lVert \kappa_{\mathrm{NN}}(x_i) - \kappa_{\mathrm{NN}}(x) \rVert \ +\\ \lVert  \kappa(x_i)\! -\! \kappa_{\mathrm{NN}}(x_i) \rVert &\leq 
        (L_{\mathrm{MPC}}\! +\!  L_{\mathrm{NN}})\lVert x\! -\! x_i \rVert + \epsilon_\mathrm{D}. \end{split}
    \end{equation*}
    Given our choice of $x_i$, \eqref{eq:delta} holds and thus
    \begin{equation*}
        \lVert \kappa(x)\!-\!\kappa_{\mathrm{NN}}(x) \rVert\ \leq\ (L_{\mathrm{MPC}} + L_{\mathrm{NN}}) \delta + \epsilon_\mathrm{D} =\epsilon.
    \end{equation*}
\end{proof}\vspace{-6pt}
As Theorem~\ref{t:1} provides a global error bound, we can view the closed-loop application of the approximate controller $\kappa_{\mathrm{NN}}(x)$ as controlling the system via the baseline MPC $\kappa(x)$ with a bounded input disturbance $e$, such that with $\lVert e \rVert \leq \epsilon$
\begin{equation}
    \begin{split}
        x(t_{n+1}) &= f(x(t_n),\kappa_{\mathrm{NN}}(x(t_n)) \\&= f(x(t_n),\kappa(x(t_n))+e). \label{eq:boundedinput}
    \end{split}
\end{equation}
This enables us to provide performance guarantees for the approximate controller $\kappa_{\mathrm{NN}}$ given that the baseline MPC is robust against bounded input uncertainties. We extend the existing results from \cite{Allgoewer2018a}, and the current work can be viewed as a constructive version of the methods presented therein. Specifically, we propose to design the training process for an arbitrary error bound $\epsilon$, so that if Assumption~\ref{a:2} holds, Theorem~\ref{t:1} replaces the need for post-verification of the error bound in \cite{Allgoewer2018a} by following the first three steps of the Algorithm~$1$ in \cite{kohler2018}. We formalize this in Corollary~1.
\begin{corollary}
    If $\kappa_{\mathrm{NN}}$ fulfills Theorem \ref{t:1} and the MPC used for data generation is robust against bounded input disturbances $e$ with $\lVert e \rVert \leq \epsilon$, then $\forall x \in \mathcal{X}$ the closed-loop in \eqref{eq:boundedinput} remains in the feasible set of the robust MPC for all times and the approximate MPC $\kappa_{\mathrm{NN}}$ is stable.
\end{corollary}
The proof of this corollary follows along the same lines as presented in \cite{Allgoewer2018a}.

\section{SENSITIVITY-REGULARIZED TRAINING}
\label{sec:sensreg}

The bound in \eqref{eq:errorbound} provides a notion of the worst-case deviation, depending on $\delta$ in~\eqref{eq:delta}. To ensure a reasonable dataset composition and training, it is intuitive that, for a given $\epsilon$, $\delta$ should be maximized to minimize the required number of data points. For example, given a state dimension $n_x$, decreasing the distance between data points by a factor $\gamma$ requires approximatly $\gamma^{n_x}$ more points. While $\epsilon$ is fixed by choice and $L_{\mathrm{MPC}}$ is fixed by the design of the underlying MPC, $\epsilon_\mathrm{D}$ and $L_{\mathrm{NN}}$ are affected by the training process. Therefore, our goal is to optimize the training procedure to reduce $\epsilon_\mathrm{D}$ and $L_{\mathrm{NN}}$ as much as possible to improve the constraints \eqref{eq:errorbound} and \eqref{eq:delta}. We propose an adaptation of the loss via the sensitivities of the OCP, which have been shown in \cite{luken2023} to improve convergence. We present a brief overview of sensitivities, followed by the adaptation of the training in \ref{subsec:lossadapt}.

\subsection{Sensitivity of Parametric Nonlinear Programs}
\label{subsec:calcsens}
For a general parametric nonlinear program of the form
\begin{mini}|s|[0]<break>
{\xi}
{c(\xi, p)}
{\label{eq:nlp}}
{}
\addConstraint{g(\xi, p) \leq 0}{,}
\end{mini}
of which the optimal control problem \eqref{eq:mpc} is a special case, the Karush-Kahn-Tucker conditions are given by
\begin{align}
    \nabla_\xi \mathcal{L}(\xi^*,\lambda^*,p) &= 0 \label{eq:kkt_stationarity}
    \\
    g(\xi^*,p) &\leq 0 \label{eq:kkt_prim_feas}
    \\
    \lambda^* &\geq 0 \label{eq:kkt_dual_feas}
    \\
    \lambda^* \odot g(\xi^*,p) &= 0 \label{eq:kkt_complementary},
\end{align}
where $\mathcal{L}(\xi,\lambda,p)=c(\xi,p)+\lambda^\top g(\xi,p)$ is the Lagrangian and $\odot$ denotes the Hadamard product.
Assuming that \eqref{eq:kkt_stationarity}{-}\eqref{eq:kkt_complementary} hold at a solution $(\xi^*,\lambda^*)$, and the Lagrangian is twice continuously differentiable near the solution, the gradient $\partial \xi^* / \partial p$ can be obtained by differentiation of said conditions \cite{fiacco1990}.
From the solution $(\xi^*,\lambda^*)$, we extract the active set and denote the active constraints as $g_\mathrm{active}$ with the corresponding Lagrange multipliers $\lambda_\mathrm{active}$, allowing us to rewrite \eqref{eq:kkt_stationarity}{-}\eqref{eq:kkt_complementary}\noeqref{eq:kkt_prim_feas}\noeqref{eq:kkt_dual_feas} as the implicit function 
\begin{align*}
    C(z^*,p) = \begin{bmatrix}
        \nabla_\xi \mathcal{L}(\xi^*,\lambda^*,p) 
        \\
        g_\mathrm{active}(\xi^*,p)
    \end{bmatrix} = 0,
\end{align*}
where $z {=} \left[ \xi;  \lambda_\mathrm{active} \right]$.
Using implicit differentiation we obtain
\begin{equation}
    \frac{\partial C}{\partial z^*} \frac{\partial z^*}{\partial p} = -\frac{\partial C}{\partial p}, \label{eq:sens-linearsystem}
\end{equation}
which can be solved for the parametric sensitivity $\partial z^* / \partial p$~\cite{andersson2018}.

\subsection{Loss Adaptation}

\label{subsec:lossadapt}
We propose an augmentation of the loss function in training by adding a sensitivity term to ensure a better fit to reduce $\epsilon_\mathrm{D}$. This results in not only minimizing the distance at each training point but also trying to match the first derivative at that point.

To learn a control law a dataset comprised of states and corresponding inputs ${\mathcal{D}=\{(x_i,\kappa(x_i))\}_{i \in \mathcal{I}}}$ is commonly generated. Then we optimize the weights of the NN via backpropagation such that the difference between the labels $\kappa(x_i)$ and the NN outputs $\kappa_{\mathrm{NN}}(x_i)$ is minimal. This difference is measured via the loss function, which is the cost function of the optimization. Here the mean squared error (MSE) is commonly used for regression tasks
\begin{equation}
    \mathcal{L}_\mathrm{MSE}(\mathcal{D},\Theta) = \frac{1}{\lvert \mathcal{I} \rvert} \sum_{i\in \mathcal{I}} \lVert \kappa(x_i) - \kappa_{\mathrm{NN},\Theta}(x_i) \rVert^2,\label{eq:MSE}
\end{equation}
where $\Theta$ are the parameters of $\kappa_{\mathrm{NN},\Theta}$.

We now also want to match the first derivative of the approximate control law $\partial \kappa_{\mathrm{NN}}(x)/\partial x$ to the first derivative of the original MPC control law $\partial \kappa(x)/\partial x$. 
The former corresponds to the derivative of the NN output with respect to its input and thus is straightforward to compute, assuming smooth activation functions, rendering the NN continuously differentiable \cite{Hastie2009}.
The latter, corresponding to the parametric sensitivity of the OCP in \eqref{eq:mpc} with respect to $x_s$, is calculated as discussed in Section~\ref{subsec:calcsens}.
With this, we extend the dataset by adding the sensitivities at each training point ${\hat{\mathcal{D}} = \{(x_i,\kappa(x_i),\partial \kappa(x_i)/\partial x_i)\}}_{i \in \mathcal{I}}$, where $\partial \kappa(x_i)/\partial x_i$ is the sensitivity at point $i \in \mathcal{I}$. In the loss function, we include an additional term to match the sensitivities
\begin{equation}
    \mathcal{L}_{\mathrm{sens}}(\hat{\mathcal{D}},\Theta)\!=\!\frac{1}{\lvert \mathcal{I} \rvert} \sum_{i\in \mathcal{I}} \left\lVert \frac{\partial \kappa(x_i)}{\partial x_i} - \frac{\partial \kappa_{\mathrm{NN},\Theta}(x_i)}{\partial x_i} \right\rVert^2\!. \label{eq:sensloss}
\end{equation}
We now add the loss terms \eqref{eq:MSE} and \eqref{eq:sensloss} weighted by $\lambda_1$ and $\lambda_2$ to create a new loss function. Furthermore, we include $\ell_2$ regularization on the weights to reduce the NN's Lipschitz constant $L_{\mathrm{NN}}$, as demonstrated in \cite{pauli2021} and weight this term by the factor $\lambda_{3}$.
\begin{equation}
    \begin{split}
        \mathcal{L}(\hat{\mathcal{D}},\Theta) =& \lambda_{3}\!\sum_{j=0}^{m} \left( \lVert W_j \rVert_{\mathrm{2}}^2 \right) + \\ &\lambda_{1}\mathcal{L}_\mathrm{MSE}(\hat{\mathcal{D}},\Theta) + \lambda_{2}\mathcal{L}_{\mathrm{sens}}(\hat{\mathcal{D}},\Theta),
    \end{split}\label{eq:fullloss}
\end{equation}
where $W_j \in \Theta$ are the layer weights as shown in \eqref{eq:nnlayer}.
The loss in \eqref{eq:fullloss} can now be optimized via regular back-propagation.

\section{Simulation Experiment}
\label{sec:sim}

\begin{figure*}
   \def\svgwidth{\linewidth}
    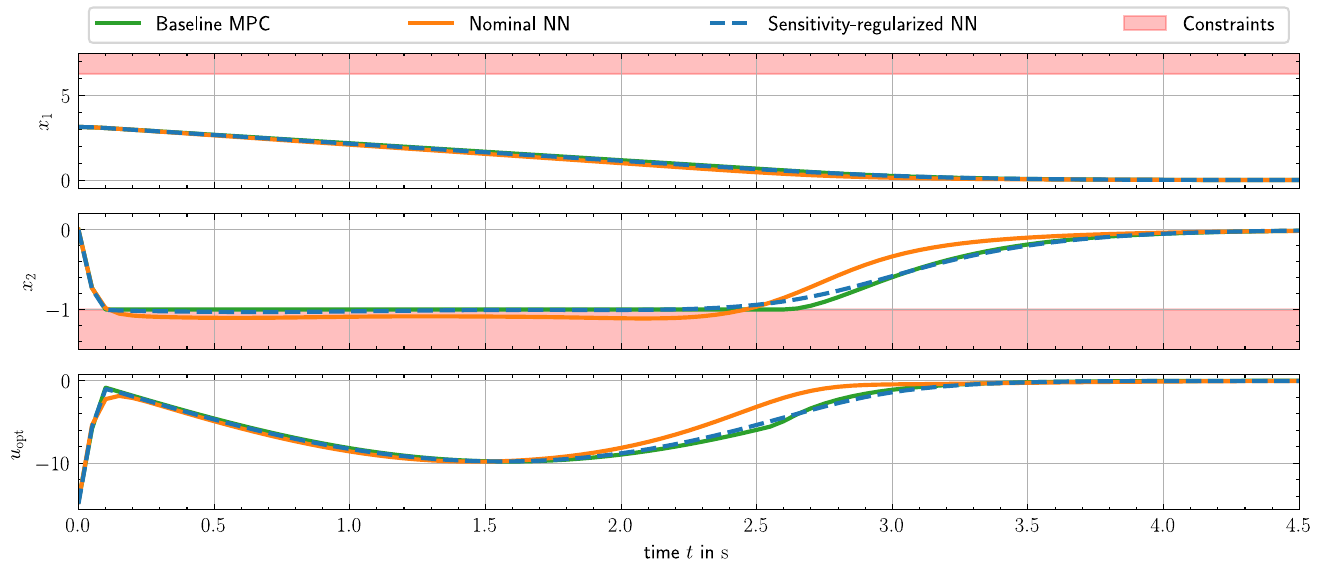
    \caption{Closed-loop simulation for $4.5~\mathrm{s}$ of the real MPC, Norminal NN, and Sensitivity-Regularized NN. Shown are the first (top) and second state (center) and the input (bottom).}
    \label{fig:closedloop}
\end{figure*}

We illustrate the proposed training method utilizing sensitivities of the optimal control problem with a simulation example of an actuated inverted pendulum with two states and one input, where both states are measured at all times
\begin{align*}
    \dot{x}_1 &= x_2,\\
    \dot{x}_2 &=  \frac{u}{ml^2} - \frac{g}{l} \sin(x_1+\pi).
\end{align*}
Here, $x_1 = \theta\!-\!\pi$, with $\theta$ being the angle of the pendulum at the downward position, $x_2 = \omega$ denoting the angular velocity, and $u=\tau$ the input torque. For parameters: $m=1~\mathrm{kg}$ is the mass of the pendulum, $g=9.81~\mathrm{m/s^2}$ is the acceleration due to gravity, and $l=1~\mathrm{m}$ is the length of the pendulum. A coordinate transformation is designed to set the upright pendulum position as the origin.\\
Then, a MPC that stabilizes the upper position of the pendulum with an optimal control problem corresponding to \eqref{eq:mpc} is designed. The cost function used is comprised of a quadratic cost ${S(x_k,u_k) = x_k^T Q x_k + u_k^T R u_k}$ and terminal cost ${V(x_N) = x_N^T P x_N}$. We chose the weights as ${Q = P = \mathrm{diag}(10,1)}$ and $R=0.1$, with $\mathrm{diag}$ denoting a diagonal matrix. The state constraints were ${-2\pi \leq x_1 \leq 2\pi}$ and ${-1 \leq x_2 \leq 1}$.\\
Using this MPC law a dataset $\hat{\mathcal{D}}$  as defined in Section~\ref{sec:sensreg} is generated by seeding $350$ equidistant grid points in $\mathcal{X}$ and solving the optimal control problem at each point, while also calculating the needed sensitivities. We trained 2 NNs with 2 hidden layers, each with $10$ neurons, \emph{tanh} activation functions, and the same initial parameters on the  dataset. A validation set of randomly chosen states in $\mathcal{X}$ was used for evaluation during training. One of the NNs was trained in the nominal scheme, only using MSE loss \eqref{eq:MSE}, and the other NN used our proposed regularization \eqref{eq:fullloss}. The NN hyper-parameters were determined via grid search for the nominal case. We trained both NNs for $2000$ epochs to ensure, that the same amount of state samples were shown to each network during training. For scaling the sensitivity-regularization loss-components, we used $(\lambda_1,\lambda_2,\lambda_3)=(1,3,0.05)$.
\begin{figure}
    \centering
   \def\svgwidth{\linewidth}
    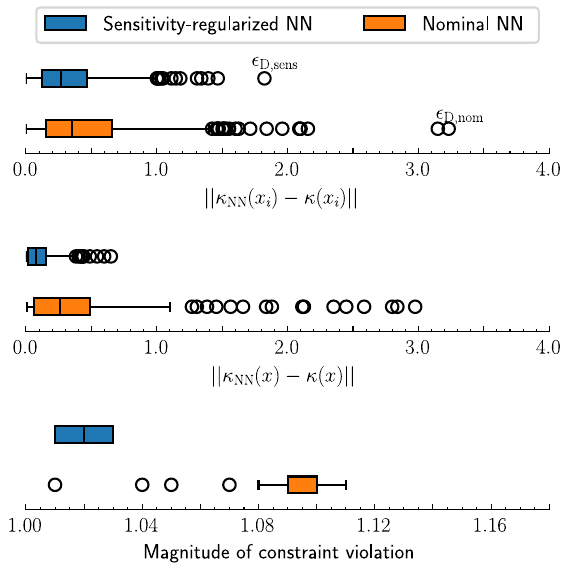
    \caption{Approximation error for the nominal NN and the sensitivity-regularized NN (top), divergence of the input values along the closed-loop trajectory (middle), and magnitude of constraint violation in closed-loop (bottom). $\epsilon_\mathrm{D}$ denoted as the maximal error according to \eqref{eq:explespd}. In all box plots, the boxes show the $25\,\%$ to $75\,\%$ quartile, and the whiskers extend to include the majority of the data.}
    \label{fig:stats}
\end{figure}
\begin{table}[!b]
    \centering
    \caption{Post-training statistics for both nominal and sensitivity-regularized neural networks.}
    \begin{tabular}{|l|r|r|r|} \hline 
         \makecell[l]{Training type}&  \makecell[l]{Validation \\ $R^2$-score} &  \textbf{$\epsilon_\mathrm{D}$}&  \makecell[l]{Lipschitz \\ constant}\\ \hlineB{2.5} 
         Nominal&  0.985&  3.231& 384.01 \\ \hline
         Sensitivity-regularized&  0.992&  1.823& 207.66\\ \hline
    \end{tabular}
    \label{tab:nnstats}
\end{table}
The NN trained with sensitivity-regularization converged faster and achieved higher validation $R^2$-score compared to the nominal NN training regiment (cf. TABLE~\ref{tab:nnstats}). After the training we evaluated the approximation error for all points in the training set, the results of this analysis are shown in Fig.~\ref{fig:stats} (top). Here it can be seen that sensitivity-regularized NN has on average lower approximation errors across all training points. Notably, in both NNs the calculated values of $\epsilon_\mathrm{D}$ are outliers, and the value of $\epsilon_\mathrm{D}$ for the sensitivity-regularized NN is significantly lower than the nominal NN's value (cf. also TABLE~\ref{tab:nnstats}). The median of the approximation errors over the dataset is also lower for the regularized NN compared to the nominal case. The sensitivity-regularized NN did furthermore converge to a lower Lipschitz constant ${L_{\mathrm{NN}} = 207.66}$ compared to the nominal NN ${L_{\mathrm{NN}} = 384.01}$.\\
The closed-loop system for the upswing of the pendulum was simulated with a simulation time of $4.5~\mathrm{s}$  for the baseline MPC and both imitation controllers, where all controllers could stabilize the system, shown in Fig.~\ref{fig:closedloop}. One can also observe a constraint violation for both NNs acting as the controllers, with the sensitivity-regularized NN showing a lower magnitude of the violation, as shown in Fig.~\ref{fig:stats} (bottom). Additionally, the sensitivity-regularized NN determines input values closer to those chosen by the MPC (cf. TABLE~\ref{tab:clstats}). We define the input divergence as ${\lVert \kappa(x) - \kappa_{\mathrm{NN}}(x) \rVert}$ along the trajectory. The statistics on the divergence are shown in Fig~\ref{fig:stats} (center).

\begin{table}[!b]
    \centering
    \caption{Statistics of closed-loop simulation for both nominal and sensitivity-regularized neural networks.}
    \begin{tabular}{|@{\ }l@{\ }|@{\ }r@{\ }|@{\ }r@{\ }|@{\ }r@{\ }|} \hline 
         \makecell[l]{Training type}&  \makecell[l]{ Constraint \\violations in $\%$} &  \makecell[l]{ Max. violation\\ magnitude}&  \makecell[l]{Max. input \\ divergence}\\ \hlineB{2.5} 
         Nominal&  46.534&  1.113& 2.975\\ \hline
         Sensitivity-regularized&  39.604&  1.032 & 0.649\\ \hline
    \end{tabular}
    \label{tab:clstats}
\end{table}

\subsection{Discussion}

The proposed regularization increases convergence speed and the prediction accuracy for our NNs, as the regularized NN reaches higher $R^2$-scores.
Furthermore, the proposed training also positively influences the $\epsilon_\mathrm{D}$ of the imitation controller and its Lipschitz constant. This leads to a less restrictive constraint in Theorem \ref{t:1}, allowing for a sparser dataset $\mathcal{D}$ to guarantee the same error bound $\epsilon$, compared to the nominally trained NN. Therefore, even if the bound presented is conservative, it provides useful guidance for training and dataset design. This increase in performance can be seen in the reduction of $\epsilon_\mathrm{D}$ and the reduced approximation error across the dataset. Notice, that the errors in the training dataset are generally larger than those along the closed-loop trajectory. This observation can be used to improve practical bounds by restricting the state domain to the combinations of states that are feasibly reachable in the closed-loop. Extending training time could help, but it risks overfitting and may increase the Lipschitz constants of the NNs.\\
We can observe that the improved open-loop prediction results translate to better closed-loop performance. The training approach results in a closer approximation of the MPC closed-loop behavior by the imitation controller. This is visible in the significant reduction of the amount and magnitude of constraint violation along the trajectory. The reduction in approximation error is particularly noticeable in the input sequence of the closed-loop. This is related to the higher frequency of low approximation errors seen in Fig. \ref{fig:stats}.\\
In summary, the presented regularization procedure increases both approximation and closed-loop performance and reduces the needed dataset density for a given error bound~$\epsilon$.

\section{CONCLUSIONS}

An approach to determine the worst-case approximation error bound was derived for a learned MPC using Lipschitz constants. This has led to a constructive algorithm to determine the required density of the training set as a function of the desired maximal approximation error. For the obtained error bound, it could be shown how stability and recursive feasibility can be proven for the imitation controller utilizing robust model predictive control.
We supplement the presented theoretical results by proposing modifications to the NN training procedure using OCP sensitivities and regularization in the cost function to reduce the needed dataset density for a specified approximation error.
In simulations, we showed that the proposed training modification reduces the needed data points, as the trained neural network achieves better error bounds on the training set and a lower Lipschitz constant. Furthermore, it was demonstrated that the approximate controller designed this way has a better closed-loop performance both in terms of matching the baseline MPC and reducing constraint violations.
In future works, we seek to improve the conditions of the underlying theorem and provide constraint guarantees by utilizing the robustness of the baseline MPC. We will further derive a tailored approach incorporating all steps from robust MPC design to the training of the imitation controller.





\bibliographystyle{IEEEtran}
\bibliography{literature}

\end{document}